\documentclass[acmtog]{acmart}

\AtBeginDocument{%
  \providecommand\BibTeX{{%
    \normalfont B\kern-0.5em{\scshape i\kern-0.25em b}\kern-0.8em\TeX}}}

\usepackage{amsmath,amsfonts}
\usepackage{algorithmic}
\usepackage{graphicx}
\usepackage{graphicx}
\usepackage{subfigure}
\usepackage{multirow}
\usepackage{enumitem}
\usepackage{cleveref}
\usepackage{pifont}
\usepackage{mathtools}
\usepackage[linesnumbered,ruled,vlined]{algorithm2e}

\copyrightyear{2022} 
\acmYear{2022} 
\setcopyright{acmcopyright}
\acmConference[NANOARCH '22]{17th IEEE/ACM International Symposium on Nanoscale Architectures}{December 7--9, 2022}{Virtual, OR, USA}
\acmBooktitle{17th IEEE/ACM International Symposium on Nanoscale Architectures (NANOARCH '22), December 7--9, 2022, Virtual, OR, USA}
\acmPrice{15.00}
\acmDOI{10.1145/3565478.3572544}
\acmISBN{978-1-4503-9938-8/22/12}

\begin{document}

\newcommand{\crl}{CryptoLight} 

\title{\crl: An Electro-Optical Accelerator for Fully Homomorphic Encryption}

\author{Mengxin Zheng}
\affiliation{%
  \institution{Indiana University of Bloomington}
  \city{Bloomington}
  \country{USA}}
\email{zhengme@iu.edu}

\author{Qian Lou}
\affiliation{%
  \institution{University of Central Florida}
  \city{Orlando}
  \country{USA}}
\email{qian.lou@ucf.edu}

\author{Fan Chen}
\affiliation{%
  \institution{Indiana University of Bloomington}
  \city{Bloomington}
  \country{USA}}
\email{fc7@iu.edu}

\author{Lei Jiang}
\affiliation{%
  \institution{Indiana University of Bloomington}
  \city{Bloomington}
  \country{USA}}
\email{jiang60@iu.edu}

\author{Yongxin Zhu}
\affiliation{%
  \institution{Shanghai Advanced Research Institute, Chinese Academy of Sciences}
  \city{Shanghai}
  \country{China}}
\email{zhuyongxin@sari.ac.cn}

\begin{abstract}
Fully homomorphic encryption (FHE) protects data privacy in cloud computing by enabling computations to directly occur on ciphertexts. To improve the time-consuming FHE operations,  we present an electro-optical (EO) FHE accelerator, \crl. Compared to prior FHE accelerators, on average, CryptoLight reduces the latency of various FHE applications by >$94.4\%$ and the energy consumption by >$95\%$.
\end{abstract}

\ccsdesc[500]{Security and privacy~Privacy protections}
\ccsdesc[500]{Hardware~Emerging optical and photonic technologies}


\maketitle

\begin{figure*}[t!]
\centering
\includegraphics[width=\linewidth]{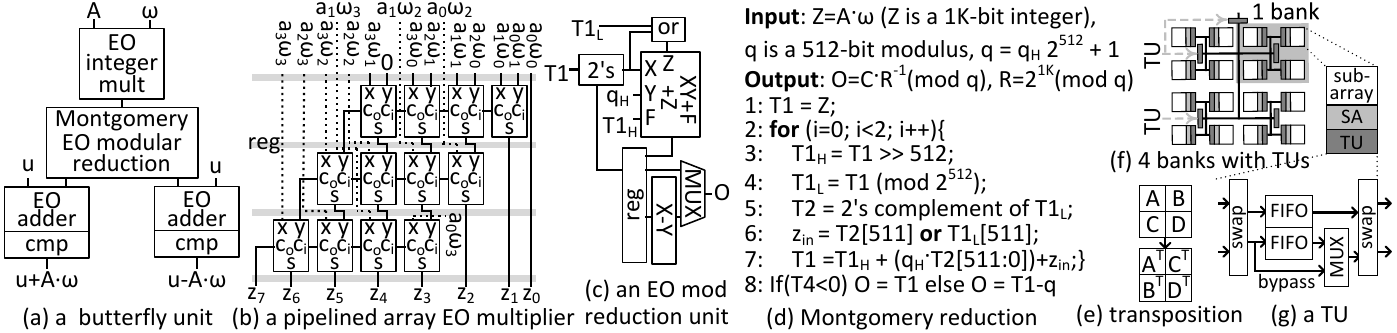}
\vspace{-0.3in}
\caption{An NTT unit in an EO CU, and a TU in a SPM.}
\label{f:photo_fhe_nttb}
\vspace{-0.1in}
\end{figure*}

\section{Introduction}
\label{s:intro}

\textit{Data privacy} is gaining tremendous importance around the world. This leads a surge in demand for privacy-preserving computing solutions protecting data confidentiality while in transit, rest, and in-use. 
\textit{Fully homomorphic encryption} (FHE)~\cite{Palisade:GPU2020} emerges as one of the most promising solutions to guaranteeing data privacy by allowing computations to directly happen on ciphertexts. 
FHE operations are extremely time-consuming, i.e., one FHE bootstrapping costs several seconds on a CPU. 
Prior work proposes GPU~\cite{Palisade:GPU2020}-, FPGA~\cite{Riazi:ASPLOS2020}-, and ASIC~\cite{Samardzic:ISCA2022,Kim:ISCA2022}-based accelerators to process FHE operations. Among all, the ASIC-based FHE accelerators, CraterLake~\cite{Samardzic:ISCA2022} and BTS~\cite{Kim:ISCA2022}, obtain the state-of-the-art performance. 

However, \textit{their performance is seriously limited by their narrow datapaths and intensive matrix transpositions}. An FHE ciphertext consists of two polynomials of large degrees (e.g., several thousand) with large integer coefficients (e.g., several hundred bits). To efficiently compute with polynomials, FHE schemes (e.g., CKKS~\cite{Samardzic:ISCA2022}) adopt Residue Number System (RNS)~\cite{Samardzic:ISCA2022}, and Number Theoretic Transform (NTT). First, to compute with large integer coefficients, RNS divides each coefficient into multiple smaller bit-width (e.g., 60-bit) residues, each of which can be processed by the datapath of prior FHE accelerators. The latency of an FHE multiplication, rotation, or bootstrapping is dominated by expensive key-switching (KS) primitives~\cite{Palisade:GPU2020} that makes output ciphertexts to be encrypted by the same secret key as the input ciphertext(s). \textit{The computational overhead of KS greatly increases with an enlarging number of residues}. Second, for two large degree-$N$ polynomials, NTT and inverse NTT (INTT) reduces the time complexity of their multiplications to $\mathcal{O}(N\log N)$. But it is difficult to perform an (i)NTT, i.e., NTT or iNTT, on a large degree-$N$ polynomial directly. Prior FHE accelerators~\cite{Samardzic:ISCA2022,Kim:ISCA2022} place it as an $n\times n$ matrix, where $N=n^2$, perform an (i)NTT on each row, multiply the matrix with some constants, transpose the matrix, and perform an (i)NTT on each row again. As a result, \textit{frequent matrix transpositions greatly prolong the latency of KS in various FHE operations by introducing huge volumes of on-chip memory traffic}.

We propose an electro-optical (EO) FHE accelerator, \textit{\crl}, to support a large bit-width datapath and free its computing units from matrix transpositions. Our contribution is summarized as:
\begin{itemize}[leftmargin=*, topsep=0pt, partopsep=0pt]
\item \textbf{A 512-bit EO CU}. We propose a 512-bit EO Computing Units (CUs) built upon ultra-fast EO integer adders and multipliers to process polynomials with large coefficients.

\item \textbf{An in-SPM TU}. We build a low-power eDRAM-based on-chip scratchpad (SPM) system.
\vspace{-0.1in}
\end{itemize}

\section{CryptoLight}
\label{s:secret}

\subsection{A 512-bit EO CU}

We build a 512-bit EO CU featured by an NTT unit, a modular add/mult unit, an automorphism unit, and a  true random number generation (TRNG) unit. Its most important component is the 512-bit EO NTT unit, which has an arithmetic and inversion unit, an address generation unit, and two butterfly units. A 512-bit EO NTT unit also supports the kernel of iNTT working in a different data flow. We also use EO adders and multipliers to construct the other CU components.

\textbf{An EO NTT Unit}. We present an EO NTT unit for the CU. Matrix transpositions are done by TUs in SPM banks, but the other three steps of the NTT on a large polynomial are computed by an NTT unit. {\crl} aims to support 64K-element NTTs, so a NTT unit supports 256-element NTT operations. The details are summarized as follows.
\begin{itemize}[leftmargin=*, topsep=0pt, partopsep=0pt, itemsep=0pt]


\item \textbf{A butterfly unit}: We propose an EO butterfly unit (BU) to accelerate radix-2 NTT butterflies, as shown in Figure~\ref{f:photo_fhe_nttb}(a). A BU consists of an EO pipelined integer array multiplier, an EO Montgomery modular reduction unit, and two EO modular adders. The EO multiplier computes the multiplication between the input and a twiddle factor $\omega$. The EO Montgomery modular reduction unit performs modular reduction on the multiplication result. By an EO adder and a comparator, the EO modular adder performs modular additions and subtractions. Two EO modular adders can generate the radix-2 butterfly outputs concurrently. 


\item \textbf{A pipelined EO array multiplier}: Through EO ripple-carry adders~\cite{Ying:JSTQE2018}, we propose an EO pipelined integer array multiplier. We show the example of a 4-bit $\times$ 4-bit pipelined array multiplier in Figure~\ref{f:photo_fhe_nttb}(b). An $m$-bit $\times$ $m$-bit pipelined array multiplier consists of $m$ stages, each of which is an $m$-bit EO ripple-carry adder. Between two stages, there is an $m$-bit register file to buffer the intermediate result. The inputs of the first pipeline stage of the multiplier are the results of AND operations between the corresponding bits of the multiplier inputs. 

\item \textbf{A Montgomery modular reduction unit}: As Figure~\ref{f:photo_fhe_nttb}(c) shows, we build an EO Montgomery modular reduction unit (MMRU) in a BU to perform modular reduction operations. The MMRU implements the modular reduction algorithm~\cite{Riazi:ASPLOS2020} shown in Figure~\ref{f:photo_fhe_nttb}(d). Besides some logic operations, and 2's complement conversions, the most intensive operation in a modular reduction operation is the multiply-add operation (i.e., $T1_{H}+(q_H\cdot T2)+z_{in}$), which can be computed by EO adders and an EO multiplier. The output of each iteration of the loop can be cached in a register file and used as the input for the next iteration. A MUX selects one between the outputs of the register file and an EO adder as the modular reduction output. 
\end{itemize}




%

\vspace{-0.15in}

\subsection{An eDRAM-based SPM System with TUs}


\textbf{Transposing a Matrix}. Matrix transpositions are heavily used in the (i)NTT and automorphism kernels of FHE operations. As Figure~\ref{f:photo_fhe_nttb}(e) shows, we implement the recursive algorithm~\cite{Samardzic:ISCA2022} to transpose a large matrix. For the transposition of a large $E\times E$ matrix, we divide the matrix into four $\frac{E}{2} \times \frac{E}{2}$ matrices, i.e., $A$, $B$, $C$, and $D$, at the top level. Instead of transposing the matrix directly, we compute $A^T$,  $C^T$, $B^T$, and $D^T$. By repeating this process recursively, the $E\times E$ matrix can be transposed. 

\textbf{In-SPM Transpose Unit}. We create an in-SPM transpose unit (TU) to transpose a matrix inside SPM banks without sending it to CUs via NoCs. As Figure~\ref{f:photo_fhe_nttb}(f) shows, along the H-tree of all SPM banks, we hierarchically deploy TUs in all SPM banks. We assume a SPM bank has four sub-arrays. Each sub-array in a SPM bank has a TU. One bank has a level-2 TU to enable data movements inside the bank. All banks share a level-3 TU for inter-bank communication. The level-3 and level-2 TUs perform the recursive matrix transposition algorithm. When the recursive process reaches the $2\times2$ matrix level, a TU attached to a sub-array swaps each element into its new position. A TUs share a similar structure shown in Figure~\ref{f:photo_fhe_nttb}(g) with different numbers of FIFOs.

\begin{figure}[ht!]
\vspace{-0.1in}
\centering
\subfigure[Latency.]{
\includegraphics[width=1.55in]{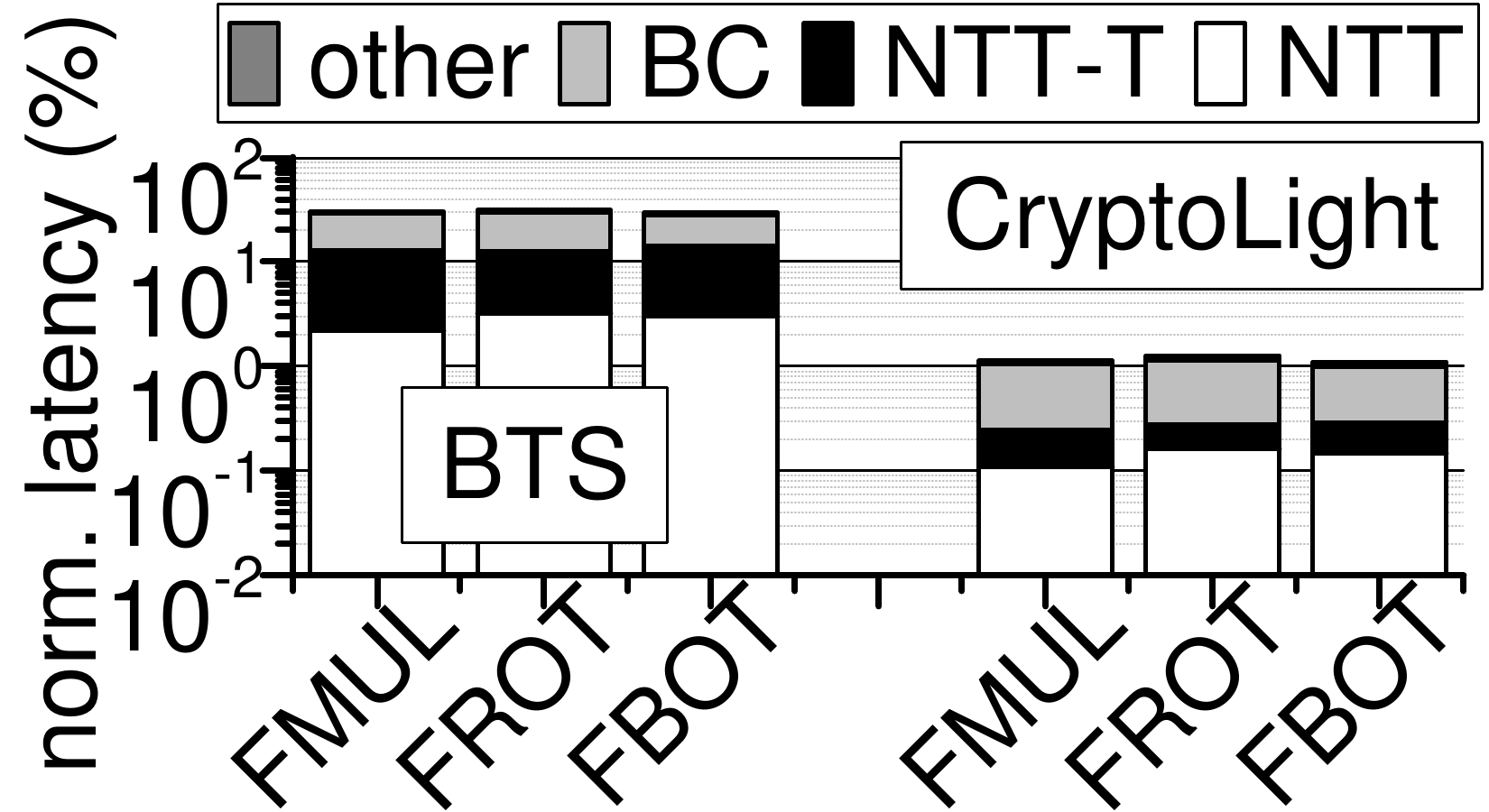}
\label{f:photo_fmul_latency}
}
\subfigure[Energy.]{
\includegraphics[width=1.55in]{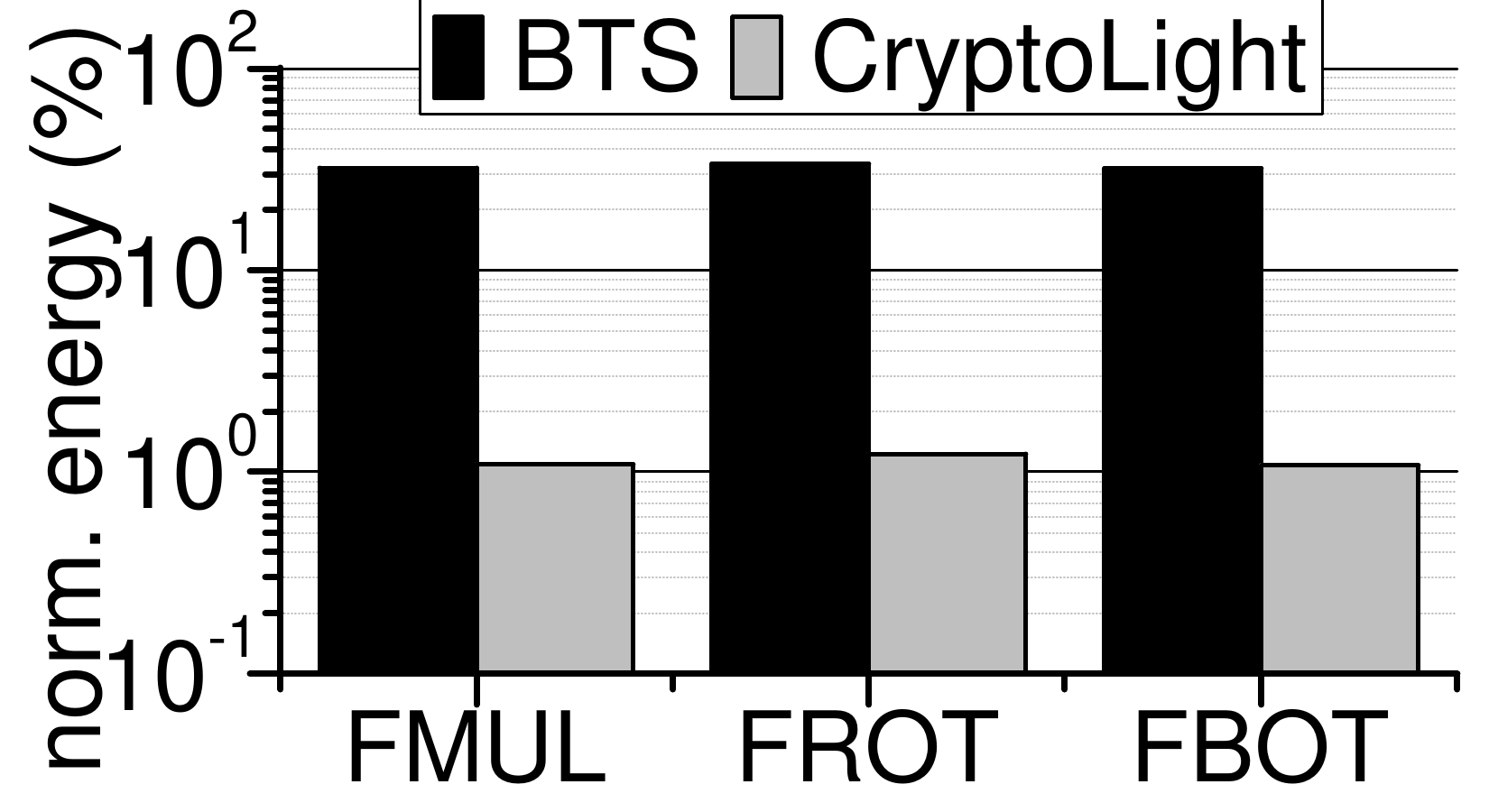}
\label{f:photo_fmul_energy}
}
\vspace{-0.2in}
\caption{The comparison of  three accelerators (norm. to Lake).}
\vspace{-0.2in}
\label{f:photo_cl_result2}
\end{figure}

\vspace{-0.1in}
\section{Experimental Methodology And Results}
\label{s:method}

 We modeled CryptoLight by a cycle-accurate FHE accelerator simulator, Sapphire-Sim~\cite{Banerjee:TCHES2019}, which is validated against several crypto-processor chips. 
 We compared CryptoLight against the state-of-the-art ASIC-based FHE hardware accelerators, CrateLake (Lake)~\cite{Samardzic:ISCA2022} and BTS~\cite{Kim:ISCA2022}. We studied the performance and energy of CKKS FMUL, FROT, and FBOT operations on ciphertexts. 

\label{s:eva}

\textbf{FMUL/FROT/FBOT latency}. The latency comparison between CryptoLight and various accelerator baselines is shown in Figure~\ref{f:photo_fmul_latency}. Compared to Lake, BTS decreases the latency of FMUL, FROT and FBOT by 69\% on average, due to its larger bit-width datapath and larger SPM system. Because of the 512-bit EO datapath and the TUs in the SPM, CryptoLight reduces the latency of FMUL, FROT and FBOT by 96\% on average over BTS.

\textbf{FMUL/FROT/FBOT energy}. The energy comparison between CryptoLight and various accelerator baselines is shown in Figure~\ref{f:photo_fmul_energy}. BTS consumes slightly larger power than Lake. Compared to BTS, CryptoLight reduces the energy consumption of FMUL, FROT and FBOT by 98.8\% on average. 

\bibliographystyle{short}
\bibliography{homo}

\begin{thebibliography}{1}
\newcommand{\enquote}[1]{``#1''}
\providecommand{\url}[1]{\texttt{#1}}
\providecommand{\urlprefix}{}

\bibitem{Banerjee:TCHES2019}
U.~Banerjee, \emph{et~al.}, \enquote{Sapphire: A Configurable Crypto-Processor
  for Post-Quantum Lattice-based Protocols,} \emph{IACR Transactions on
  Cryptographic Hardware and Embedded Systems}, (4):17–61, Aug. 2019.

\bibitem{Kim:ISCA2022}
S.~Kim, \emph{et~al.}, \enquote{BTS: An Accelerator for Bootstrappable Fully
  Homomorphic Encryption,} in \emph{ACM International Symposium on Computer
  Architecture}, 2022.

\bibitem{Palisade:GPU2020}
Palisade, \enquote{{PALISADE Operations using CUDA},} , 2020.

\bibitem{Riazi:ASPLOS2020}
M.~S. Riazi, \emph{et~al.}, \enquote{{HEAX: An Architecture for Computing on
  Encrypted Data},} in \emph{Architectural Support for Programming Languages
  and Operating Systems}, 2020.

\bibitem{Samardzic:ISCA2022}
N.~Samardzic, \emph{et~al.}, \enquote{CraterLake: A Hardware Accelerator for
  Efficient Unbounded Computation on Encrypted Data,} in \emph{IEEE/ACM
  International Symposium on Computer Architecture}, 2022.

\bibitem{Ying:JSTQE2018}
Z.~Ying, \emph{et~al.}, \enquote{{Electro-Optic Ripple-Carry Adder in
  Integrated Silicon Photonics for Optical Computing},} \emph{Journal of
  Selected Topics in Quantum Electronics}, 2018.

\end{thebibliography}
\end{document}